\documentclass[12pt,pdflatex]{article}
\usepackage{amssymb,amsmath,amsfonts,amsthm,graphicx}
\def\EEE{E_1{}^1}

\def\be{\begin{equation}}
\def\ee{\end{equation}}

\def\sech{\text{sech}}
\def\Sm{\Sigma_-}
\def\Sc{\Sigma_\times}
\def\Nm{N_-}
\def\Nc{N_\times}
\def\Sp{\Sigma_+}

\def\dT{\partial_T}
\def\dX{\partial_X}
\def\e{{\rm e}}
\def\arctanh{{\rm arctanh}}

\begin{document}

\begin{center}
{\Large\bf General relativistic density perturbations}
\vspace{.3in} \\
{\bf W C Lim},
\\Department of Mathematics, University of Waikato, Private Bag 3105, Hamilton 3240, New Zealand
\\Email: wclim@waikato.ac.nz
\vspace{.1in}
\\{\bf A A Coley},
\\Department of Mathematics \& Statistics, Dalhousie University,\\
Halifax, Nova Scotia, Canada B3H 3J5
\\Email: aac@mathstat.dal.ca
\vspace{.1in}
\vspace{0.2in}

\end{center}

[PACS: 98.80.Jk]

\begin{abstract}
We investigate a general relativistic mechanism in which 
spikes generate matter overdensities in the early universe.
When the cosmological fluid is tilted, the tilt provides another mechanism in generating matter
inhomogeneities.  
We numerically investigate the effect of a sign change in the tilt,
when there is a spike but the tilt does not change sign,
and when the spike and the sign change in the tilt coincide. 
We find that the tilt plays
the primary role in generating matter inhomogeneities, and it does so by creating
both local overdensities and underdensities.
We discuss of the physical implications of the work.
\end{abstract}

\section{Introduction}

As the first step in our effort to explore general relativistic mechanisms that
cause spacetime and matter inhomogeneities, in \cite{art:ColeyLim2012} we
concentrated on how spikes generate matter overdensities in a radiation fluid in a
special class of inhomogeneous models.  These spikes occur in the initial
oscillatory regime of general cosmological models.  The mechanism of spike formation
is simple -- the state-space orbits of nearby worldlines approach a saddle point; if
this collection of orbits straddle the stable manifold of the saddle point, then one
of the orbits becomes stuck on the stable manifold of the saddle point and heads
towards the saddle point, while the neighbouring orbits leave the saddle point.
This heuristic argument holds as long as spatial derivative terms have negligible
effect.  In the case of spikes, the spatial derivative terms do have a significant
effect, and the spike point that initially got stuck does leave the saddle point
eventually, and the spike that formed becomes smooth again.  
In the initial oscillatory regime, spikes recur~\cite{art:Anderssonetal2005,art:Lim2008,art:Limetal2009}.

With a tilted fluid, the tilt provides another mechanism in generating matter
inhomogeneities through the divergence term (in the continuity equation).  We will discover that the tilt plays
the primary role in generating matter inhomogeneities, but it does so by creating
local overdensities and underdensities without particular preference for either one.
On the other hand, the spike mechanism plays a secondary role in generating matter
inhomogeneities -- it drives the divergence term towards making local overdensities.

In this paper we first present the evolution equations of 
the orthogonally transitive $G_2$ model with a perfect fluid, which we will use in the 
numerical simulations. We briefly review the dynamics of tilt transitions.
We then numerically investigate the effect of a sign change in the tilt, which 
leads to both overdensities and undersities through a large divergence
term.
We then determine that when there is a complete spike but the tilt does not change sign,
the spikes leave a negligible imprint on the matter density.
Finally, when the spike and the sign change in the tilt coincide, we find that
the spike drives the divergence term towards making overdensities as the universe expands.
We conclude that it is the tilt instability that plays the primary role in the formation of matter inhomogeneities. 
We finish with a discussion of the physical implications of the work.

\section{Equations}

The model is an orthogonally transitive $G_2$ model (also called a Gowdy model) with Killing vector fields acting on a plane, and with a perfect fluid.
In order to study the evolution numerically,
we use the zooming technique developed in \cite{art:Limetal2009} to avoid specifying boundary conditions.
The coordinate variable $T$ increases towards the singularity.

We shall omit the derivation of the evolutions equations; for a derivation involving a perfect fluid 
see \cite{thesis:Lim2004}, and
for the derivation involving the zooming technique see \cite{art:Limetal2009}.
The evolution equations (for our numerical investigation) are:
\begin{align}
        \dT\ln\beta &= - AX \dX\ln\beta + \frac32(1-\Sp) - \frac34(2-\gamma)\frac{1-v^2}{G_+}\Omega
\\
        \dT\ln\EEE &= - AX \dX\ln\EEE - 1 + \frac34(2-\gamma)\frac{1-v^2}{G_+}\Omega
\intertext{}
        \dT\Sm &= - AX \dX\Sm
                + \tfrac12\e^{AT}\EEE \dX\Nc
                + \frac34(2-\gamma)\frac{1-v^2}{G_+}\Omega \Sm - \sqrt{3}(\Sc^2-\Nm^2)
\\
        \dT\Nc &= - AX \dX\Nc
                + \tfrac12\e^{AT}\EEE \dX\Sm
                - \Nc
		+ \frac34(2-\gamma)\frac{1-v^2}{G_+}\Omega \Nc
\\
        \dT\Sc &= - AX \dX\Sc
                - \tfrac12\e^{AT}\EEE \dX\Nm
		+ \frac34(2-\gamma)\frac{1-v^2}{G_+}\Omega \Sc
                + \sqrt{3}\Sm\Sc + \sqrt{3}\Nc\Nm
\\ 
        \dT\Nm &= - AX \dX\Nm
                - \tfrac12\e^{AT}\EEE \dX\Sc
                - \Nm
		+ \frac34(2-\gamma)\frac{1-v^2}{G_+}\Omega \Nm
                - \sqrt{3}\Sm\Nm - \sqrt{3}\Sc\Nc
\\
	\dT\ln\Omega &= - AX \dX\ln\Omega 
		- \frac{\gamma v}{2G_+} \e^{AT}\EEE \dX\ln\Omega
		+ \frac{\gamma G_-(1-v^2)}{2G_+^2} \dX \arctanh v
\notag\\
		&\quad- \frac{\gamma}{G_+} \left[ \frac{G_+}{\gamma}(q+1)-\frac12(1-3\Sp)(1+v^2)-1 \right]
\\
	\dT\arctanh v &= - AX \dX\arctanh v
		+ \frac{(\gamma-1)(1-v^2)}{2 \gamma G_-} \e^{AT}\EEE \dX\ln\Omega
\notag\\        
		&\quad- [3\gamma-4-(\gamma-1)(4-\gamma)v^2] \frac{v}{2 G_+ G_-} \e^{AT}\EEE \dX\arctanh v
\notag\\
		&\quad+ \frac{1}{2\gamma G_-} \left[ (2-\gamma)G_+ r - \gamma v(3\gamma-4 + 3(2-\gamma)\Sp) \right]
\label{dTv}
\end{align}  
where $\Sp$, $q$, $r$, $G_\pm$ are given by~{\footnote{The parameter $A$ controls the zoom rate (as explained in Section III of 
\cite{art:Limetal2009}); here we just choose               
$A=1$, so that the particle horizon stays at $X=1$.
Note that in the temporal gauge chosen the Gauss and Codacci constraints reduce to algebraic 
conditions on the dynamical variables~\cite{art:vEUW2002}.}} 
\begin{align}
	\Sp &= \frac12\left( 1-\Sm^2-\Sc^2-\Nm^2-\Nc^2-\Omega\right)
\\
	q &= 2-3\Sp - \frac32(2-\gamma)\frac{1-v^2}{G_+}\Omega
\\
	r &= -3\Nc\Sm + 3\Nm\Sc - \frac{3\gamma v}{2 G_+} \Omega
\\
	G_\pm &= 1 \pm (\gamma-1) v^2.
\end{align}

In situations where shock waves develop, we use the upwind method.
For upwind method we evolve the variables in two stages using the Godunov splitting -- the first stage we evolve only the PDE part,
the second stage only the ODE part. Both parts use the timestep $\Delta T$. The second stage uses the new data obtained in the first stage as data at time $T$.
~{\footnote{We did not use any packages like CLAWPACK or RNPL; rather we wrote our own code      
in Fortran. See \cite{book:LeVeque1992} for the background on the Godunov splitting.}}

The first stage requires eigen-functions be evolved using the upwind method.
The eigen-functions are
\be
	\ln\beta,\ \ln\EEE,\ \Sm\pm\Nc,\ \Sc\pm\Nm,\ f_\pm,
\ee
with corresponding eigen-velocities
\be
	AX,\ AX,\ AX \mp \tfrac12\e^{AT}\EEE,\ AX \pm \tfrac12\e^{AT}\EEE,\ v_\pm,
\ee
where $f_\pm$ and $v_\pm$ for the perfect fluid are given by
\be
	f_\pm = \ln \left[ \frac{1-v^2}{G_+} \left(\frac{1+v}{1-v}\right)^{\pm \frac{\gamma}{2s} }\Omega \right],
\ee
\be
	v_\pm = AX - \tfrac12\e^{AT}\EEE G_-^{-1} \left[ (2-\gamma)v \pm s (1-v^2) \right],
\ee
where $s=\sqrt{\gamma-1}$.
With these variables, the PDE part of the evolution equations is rewritten as:
\begin{align}
	\dT\ln\beta &= - AX \dX\ln\beta 
\\
	\dT\ln\EEE &= - AX \dX\ln\EEE
\\
	\dT(\Sm\pm\Nc) &= (AX \mp \tfrac12\e^{AT}\EEE) (\Sm\pm\Nc)
\\
	\dT(\Sc\pm\Nm) &= (AX \pm \tfrac12\e^{AT}\EEE) (\Sc\pm\Nm)
\\
	\dT f_\pm &= v_\pm f_\pm
\end{align}

The second stage uses the new data obtained in the first stage as data at time $T$. It evaluates the ODE part of the evolution equations.
We shall use the fourth-order Runge-Kutta method.
The ODE part of the geometric quantities can be evolved using the original variables
\be
	\ln\beta,\ \ln\EEE,\ \Sm,\ \Nc,\ \Sc,\ \Nm.
\ee
Their ODE part is straightforward, for example,
\be
        \dT\Sm = \frac34(2-\gamma)\frac{1-v^2}{G_+}\Omega \Sm - \sqrt{3}(\Sc^2-\Nm^2)
\ee
The ODE part of the evolution equations for perfect fluid is rewritten as:
\begin{align}
	\dT f_+ &= -\frac1{2s(1+sv)}\Big[ 3\gamma (s+v)\Sp + 2q - (2-\gamma) (1-sv)r
\notag\\
	&\quad -\frac{1}{G_+}\left(v-2vs^4-s+s^3-3vs^2+3s^5v^2-v^3s^4+v^3s^2+v^2s^3\right)\Big]
\\
        \dT f_- &= -\frac1{2s(1-sv)}\Big[ 3\gamma (s-v)\Sp + 2q + (2-\gamma) (1+sv)r
\notag\\
        &\quad -\frac{1}{G_+}\left(-v+2vs^4-s+s^3+3vs^2+3s^5v^2+v^3s^4-v^3s^2+v^2s^3\right)\Big] 
\end{align}
where $s=\sqrt{\gamma-1}$.

\section{Dynamics}

Before we present the numerical results, we review some background on tilt transitions \cite{art:Ugglaetal2003}. The tilt variable $v$ here 
in orthogonally transitive $G_2$ models
is the only remaining nonzero component $v_1$ of the velocity of the perfect fluid relative to the observer.
$v$ is bounded by -1 and 1. $T$ increases towards big bang singularity.

\begin{figure}[t!]
\begin{center}

    \includegraphics{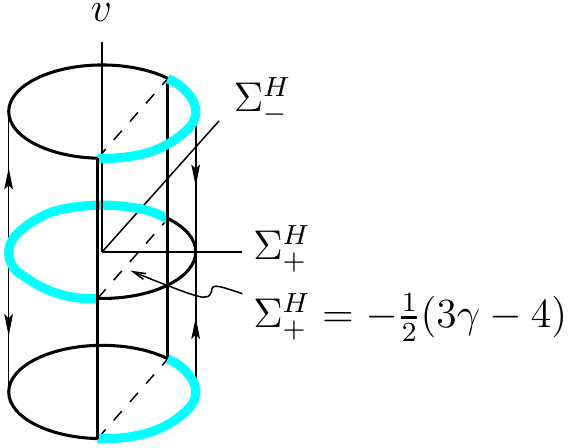}
    \caption{Tilt transitions shown in the state space. The faint arcs on the Kasner circles are unstable towards the singularity.}
    \label{tiltfig}

\end{center}
\end{figure}

Linearization of (\ref{dTv}) on the Kasner circle with zero tilt gives
\be
\label{linear_v}
	\dT v = -\tfrac12[3\gamma-4+3(2-\gamma)\Sp] v,
\ee
which gives the linear solution
\be
	v = v_0 e^{-\frac12[3\gamma-4+3(2-\gamma)\Sp]T}.
\ee
Linearization on the Kasner circle with extreme tilt $v^2=1$ gives
\be
\label{linear_v1}
	\dT (1-v^2) = \tfrac{1}{(2-\gamma)}[3\gamma-4+3(2-\gamma)\Sp] (1-v^2).
\ee
Recall that the Hubble-normalized $\Sp^H$ is given by $\Sp^H = \Sp/(1-\Sp)$.
As $T$ increases, the tilt is unstable on the arc $\Sp^H < -\tfrac12(3\gamma-4)$ of the Kasner circle with zero tilt. See Figure~\ref{tiltfig}.
We consider $1 \leq \gamma \leq 2$.
At the lower bound for $\gamma$, $\Sp^H < \tfrac12$ for dust, which makes the tilt unstable for most of the Kasner epochs (since most of the Kasner epochs are concentrated near $\Sp^H = -1$).
At the upper bound for $\gamma$, $\Sp^H < -1$ for the stiff fluid. 
But since $\Sp^H \geq -1$ by the Gauss constraint, the tilt is always stable for the stiff fluid.
For the numerical simulations in this paper, we will restrict to the physically relevant case of the radiation fluid ($\gamma=\frac43$), where the tilt is unstable for $\Sp^H < 0$. The tilt is still unstable for most of the Kasner 
epochs in this case.
We also note that growth rate near $v^2=1$ is larger than that near $v=0$, which will play a role in tendency to form shockwaves near $v^2=1$ first.
See \cite[pages 59-61]{thesis:Lim2004} for more discussion on tilt instability.

Based on the above knowledge, it is expected that, in the absence of other
transitions, an initial value $v_0(x)$ with a sign change combined with
tilt instability will develop (towards singularity) into a step function at
the position of sign change \cite[pages 143-147]{thesis:Lim2004}.

In the next three sections
we begin our investigation by studying the effect of a sign change in the tilt, in the context of diagonal $G_2$ models. We will see that the sign change in tilt makes both overdensities and undersities through large divergence
term.
We then ask the next question -- what happens when there is a spike, but the tilt does not change sign? Does the spike leave any imprint on the matter density? We will see that the answer is ``negligible".
Finally, what happens when the spike and the sign change in tilt coincide? We will see that spike drives the divergence term towards making overdensities as the universe expands.

\section{Sign change in tilt and its role through the divergence term}

We shall consider the radiation fluid hereafter.%
\footnote{Other perfect fluids with $1 < \gamma < 2$ have the same qualitative behaviour.}
$\beta$ decouples from the system and is irrelevant to the dynamics.
We begin our investigation by studying the effect of a sign change in tilt, in the context of diagonal $G_2$ models. 
Consider two numerical simulations with the following initial data (see below):
\be
\label{IC_diag}
	\Sm = - \frac{7}{2\sqrt{3}},\ \Nc = 0 = \Sc = \Nm, \ \Omega = 10^{-5},\ v = \pm\frac{X}{100},\ \EEE = 2e^{-10},
\ee
with the slope of $v$ playing an important role in deciding the sign of the divergence term $\dX v$ in the evolution of $\Omega$.

We run the simulation towards the singularity.
We use 1101 grid points on $X \in [-1.1, 1.1]$
~{\footnote{This interval was chosen to cover the dynamics inside of                         
the particle horizon of a single observer only; nothing
of interest occurs outside.}}  and run the simulations from $T=0$ to $T=20$. 
~{\footnote{Throughout the numerical evolution we used the time step size 
$\Delta T  = \Delta X /(2.2)= 0.000909$; the overall accuracy is
roughly of order ${(\Delta X)}^4 = (0.002)^4 = 1.6 \times 10^{-11}$, although in practice 
the errors were of order $10^{-10}$.}} 
The particle horizon of the observer at $X=0$ is located approximately at $X=\pm1$.

\begin{figure}[t!] 
\begin{center}
    \includegraphics[width=\linewidth]{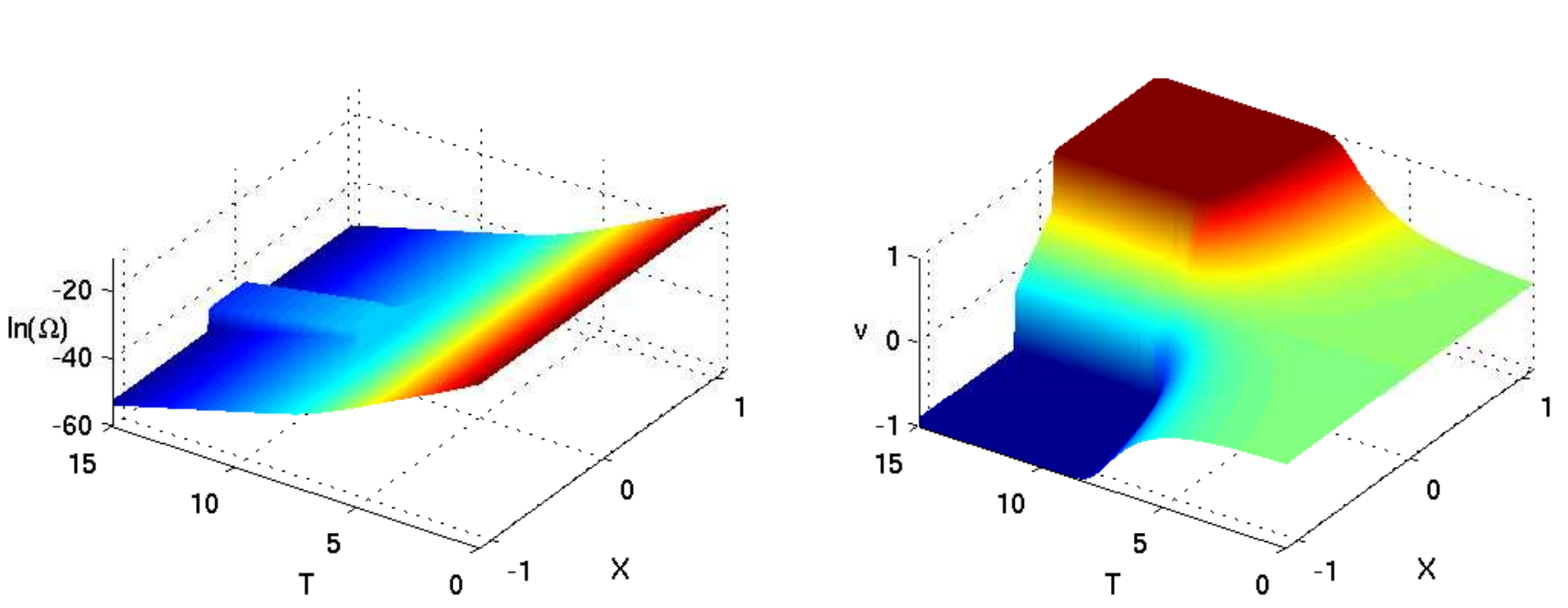}
    \caption{Plots of $\Omega$ and $v$ from the simulation with I.C.~(\ref{IC_diag}) with plus sign.}
    \label{diag_g_rad_eigen}
\end{center}
\end{figure}

\begin{figure}[t!]
\begin{center}
    \includegraphics[width=0.5\linewidth]{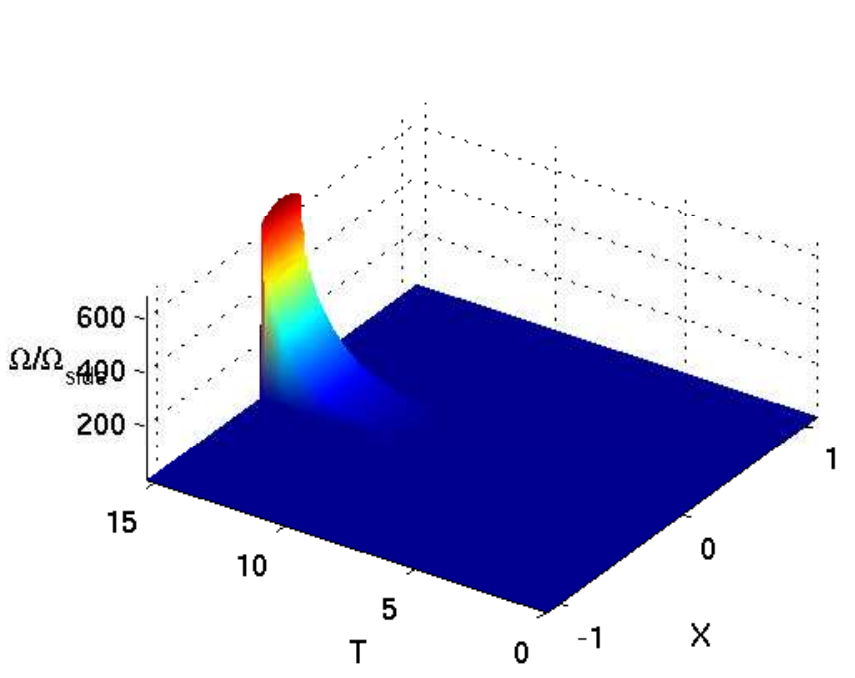}
    \caption{Corresponding ratio $\Omega/\Omega_{X=-1.1}$. $\Omega$ develops an overdensity towards singularity.}
    \label{diag_g_rad_eigen-sideo}
\end{center}
\end{figure}

The first initial data with $v = \frac{X}{100}$ gives a positive divergence term, which negates the contribution from the algebraic terms. As a result $\Omega$ develops a local overdensity towards the singularity.
See Figures~\ref{diag_g_rad_eigen} and~\ref{diag_g_rad_eigen-sideo}.
In this simulation it turns out that the divergence term even dominates the algebraic terms for a brief period and as a result $\Omega$ actually increases towards the singularity during this bried period. Two shockwaves also form.
Interpreting this result in the forward time direction (i.e. expanding away from singularity), the positive divergence term means that the fluid flows away from $x=0$, creating a local underdensity as the universe expands.

\begin{figure}[t!]
\begin{center}
    \includegraphics[width=\linewidth]{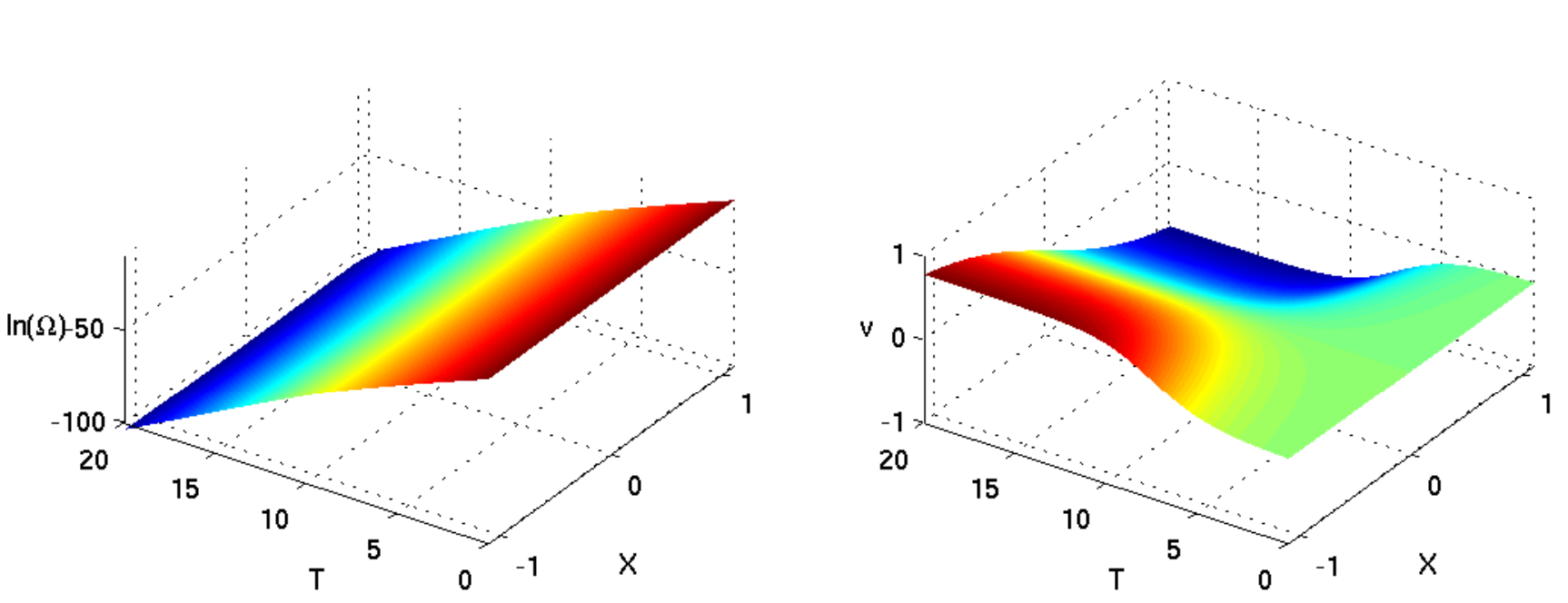}
    \caption{Plots of $\Omega$ and $v$ from the simulation with I.C.~(\ref{IC_diag}) with minus sign.}
    \label{diag_g_rad_eigen_v2}
\end{center}
\end{figure}

\begin{figure}[t!]
\begin{center}
    \includegraphics[width=0.5\linewidth]{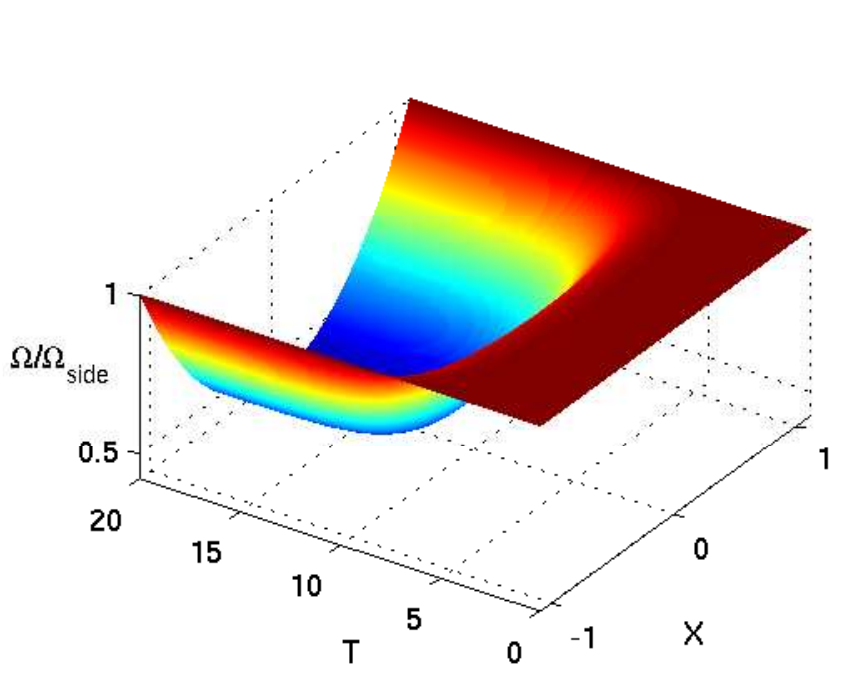}
    \caption{Corresponding ratio $\Omega/\Omega_{X=-1.1}$. $\Omega$ develops an underdensity towards singularity.}
    \label{diag_g_rad_eigen_v2-sideo}
\end{center}
\end{figure}

The second initial data with $v = -\frac{X}{100}$ gives a negative divergence term, which adds to the contribution from the algebraic terms. As a result $\Omega$ develops a local underdensity towards the singularity.
See Figures~\ref{diag_g_rad_eigen_v2} and~\ref{diag_g_rad_eigen_v2-sideo}.
Interpreting this result in the forward time direction, the negative divergence term means that the fluid flows towards $x=0$, creating a local overdensity as the universe expands.

So we see that the divergence term, depending on its sign, can create a local underdensity or overdensity. The tilt instability in the oscillatory regime makes the divergence term larger at surfaces where the tilt 
changes sign. Also, at shock surfaces, the divergence term is very large. Both $\Omega$ and $v$ develop shockwaves.

The two shockwaves form at surfaces where $v^2$ tends to 1 quickly. From the linearized evolution equations (\ref{linear_v}) and (\ref{linear_v1}) we see that $v^2$ tends to 1 at a faster rate than $v$ is leaving 0. This leads to 
the tendency to form shockwaves near $v^2=1$ first.

\section{Combining tilt transitions with other transitions}

We now shift our attention to orthogonally transitive $G_2$ models.  Here $\Sc$ and
$\Nm$ are not zero.  $\Nm$ is the trigger for a Bianchi Type II curvature transition;
$\Sc$ for a frame transition.  A sign change in $\Nm$ is responsible for a spike
transition (and sign change in $\Sc$ for a false spike transition).  See
\cite{art:Lim2008} for exact solutions and state space diagrams.  We focus on the
physical transitions, namely the Bianchi Type II transition and the spike
transition.  How large of a matter inhomogeneity can they generate?  To isolate their
effect, we study the case without a sign change in the tilt.  We use a perturbed spike
solution as the initial data

\be
\label{IC2}
	(\Sm,\ \Nc,\ \Sc,\ \Nm) = (-c \Sm{}_\text{Taub}-\tfrac{1}{\sqrt{3}},\ s \Nm{}_\text{Taub},\ c \Nm{}_\text{Taub},\ -s \Sm{}_\text{Taub}),\ \Omega = 10^{-5},\ v = \tanh(10^{-2}),
\ee
where
\be
	\Sm{}_\text{Taub} = \tfrac{1}{\sqrt{3}}[\tanh(w(T-T_0))-1],\ \Nm{}_\text{Taub} = \tfrac{w}{\sqrt{3}}\sech(w(T-T_0)),
\ee
\be
	c = \frac{f^2-1}{f^2+1},\ s = \frac{2f}{f^2+1},\ f =  w e^{T-T_0} \sech(w(T-T_0)) \frac{2}{\EEE} (X-X_0).
\ee
We choose the following parameters: $w=1.5$, so that $\Sm$ is close to the value $ - \frac{7}{2\sqrt{3}}$ in (\ref{IC_diag});
we choose $T=0$, $T_0=-10$ so that the spike transition occurs at $T=10$; we choose $\EEE = 2 e^{T-T_0}$ to cancel out these factors in $f$;
and lastly we choose to zoom in on $X_0=1$ (away from the spike) and $X_0=0$ (at the spike) in two separate simulations. 
We use 1101 grid points on $X-X_0 \in [-1.1, 1.1]$ and run the simulations from $T=0$ to $T=20$. The particle horizon of the observer at $X=X_0$ is located approximately at $X-X_0=\pm1$.
The choice $v = 10^{-2}$ above is made to avoid a sign change in the tilt, and indeed in the simulations the sign of $v$ remains positive and $v$ tends to $1$.

\begin{figure}[t!]
\begin{center}
    \includegraphics[width=\linewidth]{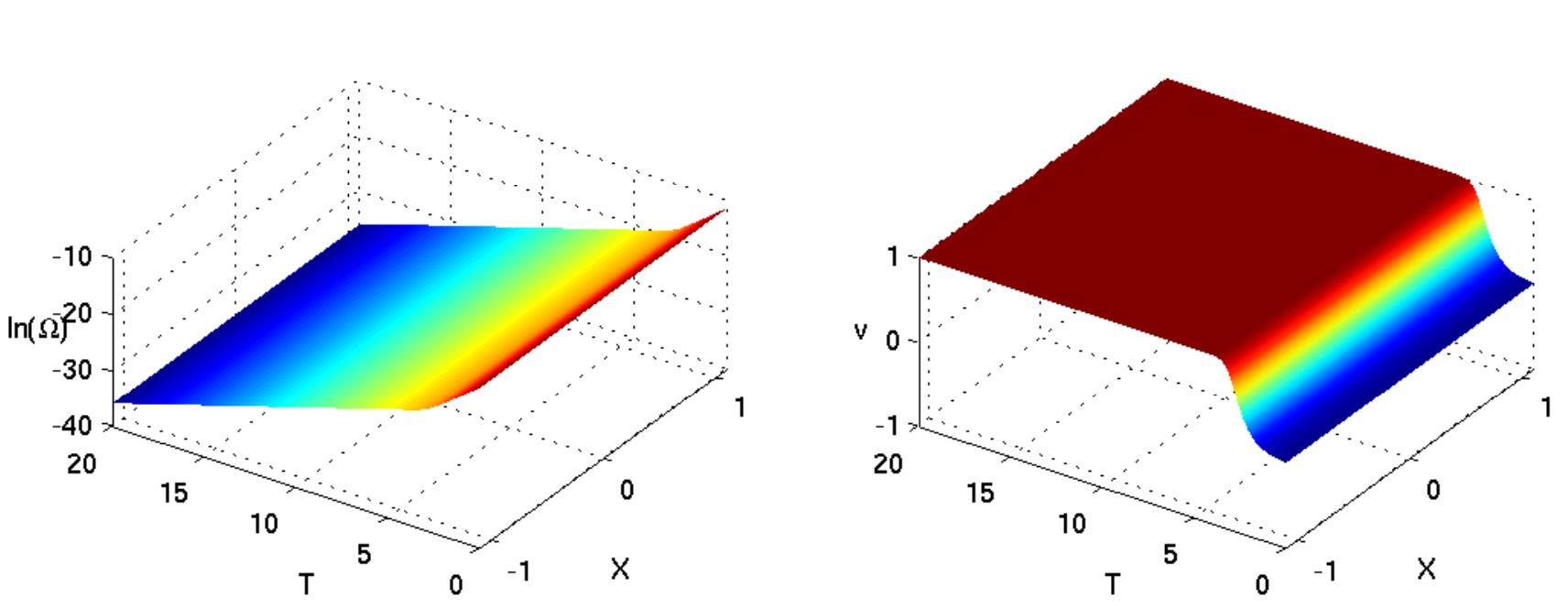}
    \caption{Plots of $\Omega$ and $v$ from the simulation with I.C.~(\ref{IC2}), zooming in at $x=0$.}
    \label{g_rad_positive}
\end{center}
\end{figure}

Zooming in at $X_0=0$, the dynamics here is simple -- $v$ tends to 1 during $T \in
[0,4]$, and then a spike transition occurs during $T \in [6,14]$.  $\Omega$ does not
develop any observable sub-horizon inhomogeneities.  Its value at $T=20$ is $2.748 \times
10^{-16}$.  See Figure~\ref{g_rad_positive}.

\begin{figure}[t!]
\begin{center}
    \includegraphics[width=\linewidth]{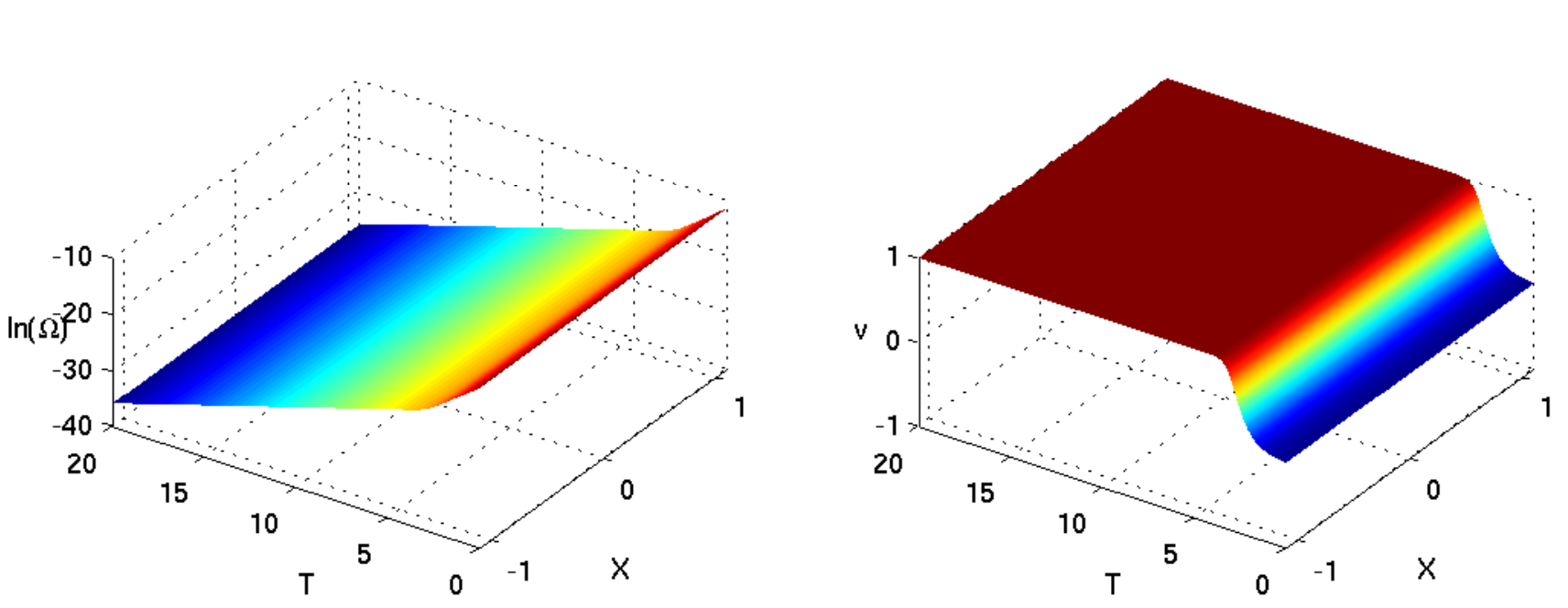}
    \caption{Plots of $\Omega$ and $v$ from the simulation with I.C.~(\ref{IC2}), zooming in at $x=1$.}
    \label{g_rad_positive_X1}
\end{center}
\end{figure}

Zooming in at $X_0=1$, the dynamics here is simpler and more homogeneous -- $v$
tends to 1 during $T \in [0,4]$, and then a curvature transition and a frame
transition occur.  The second curvature transition has not occurred before the end
of simulation, though its trigger $\Nm$ is growing slowly.  $\Omega$ does not
develop any observable sub-horizon inhomogeneities.  Its value at $T=20$ is $2.749 \times
10^{-16}$.  See Figure~\ref{g_rad_positive_X1}, which is virtually identical to
Figure~\ref{g_rad_positive}.

\begin{figure}[t!]
\begin{center}
    \includegraphics[width=\linewidth]{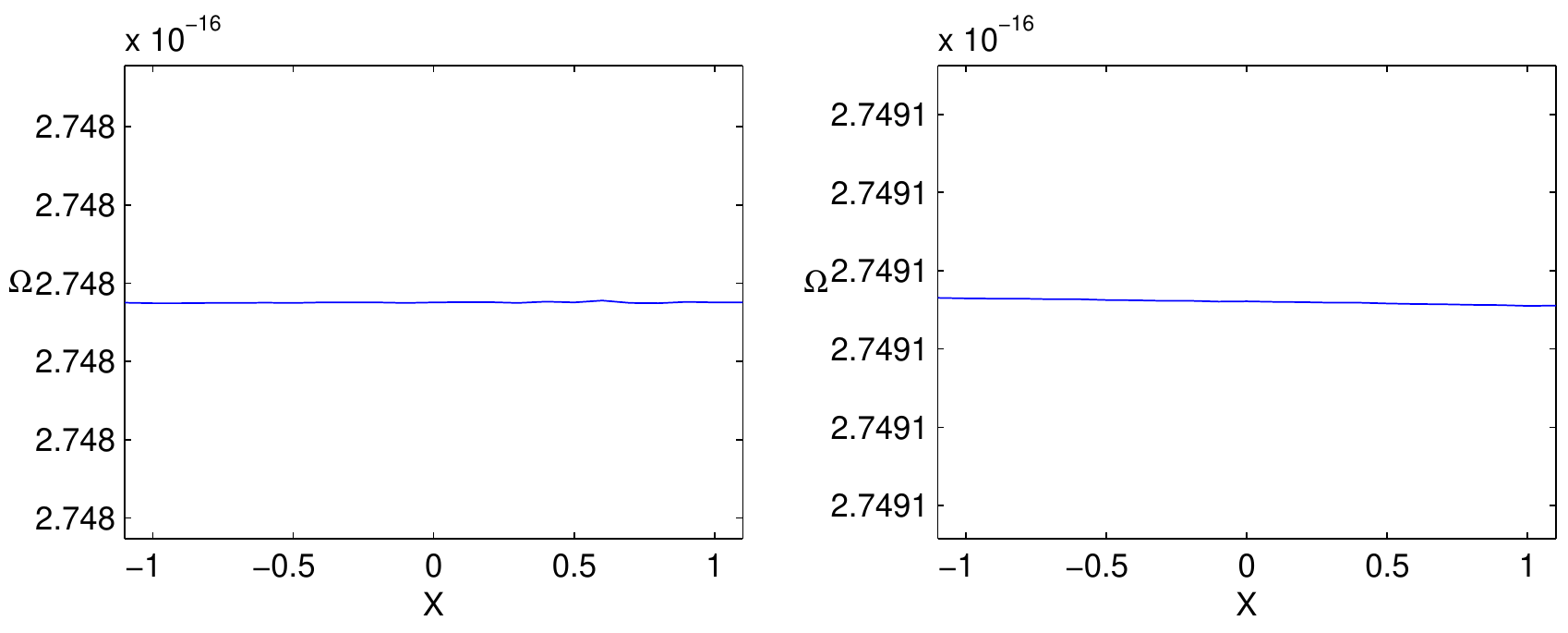}
    \caption{$\Omega$ at $T=20$ for zoom-in at $x=0$ (left panel) and $x=1$, showing that spike alone leaves virtually no imprint on matter.}
    \label{g_rad_positive-Omegas}
\end{center}
\end{figure}

The plots of $\Omega$ at $T=20$ for the two simulations
are compared in Figure~\ref{g_rad_positive-Omegas}.
We conclude that the spike transition and the curvature transition do not any create observable
sub-horizon inhomogeneities, at least when $v^2$ is close to 1.

\section{Combining sign change in tilt and spike transitions}

We now know that a sign change in the tilt generates matter inhomogeneities through the divergence term. But the divergence term can be positive or negative, and has no preference.
Do spike transitions influence this preference?
To see this, we combine a sign change in the tilt with these transitions.
We use the same initial data in (\ref{IC2}) except for $v$.
Here we use $v = -\tanh(X/100)$, so that initially the divergence term is negative. We zoom in at $X_0=0$.
We use 1101 grid points on $X-X_0 \in [-1.1, 1.1]$ and run the simulations from $T=0$ to $T=20$. The particle horizon of the observer at $X=X_0$ is located approximately at $X-X_0=\pm1$.

\begin{figure}[t!]
\begin{center}
    \includegraphics[width=\linewidth]{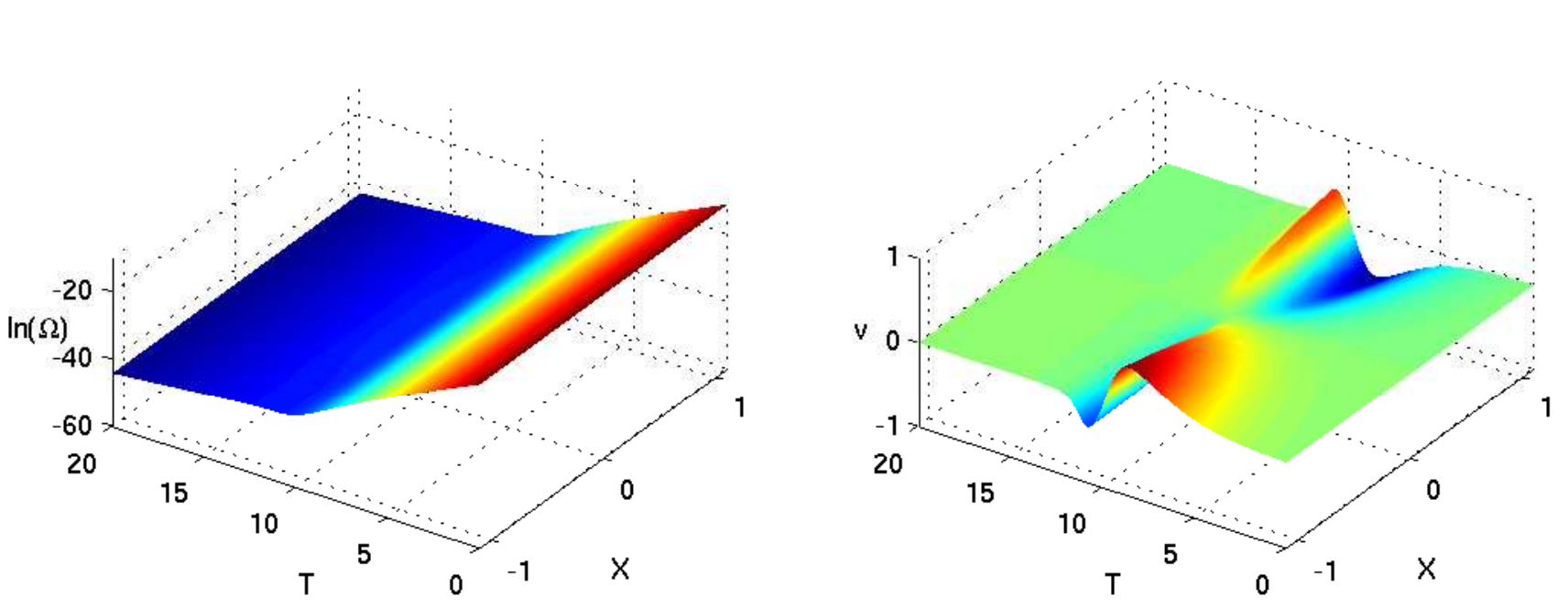}
    \caption{Plots of $\Omega$ and $v$ from the simulation with I.C.~(\ref{IC2}) but with $v=\tanh(-X/100)$.}
    \label{g_rad_v3}
\end{center}
\end{figure}

\begin{figure}[t!]
\begin{center}
    \includegraphics[width=\linewidth]{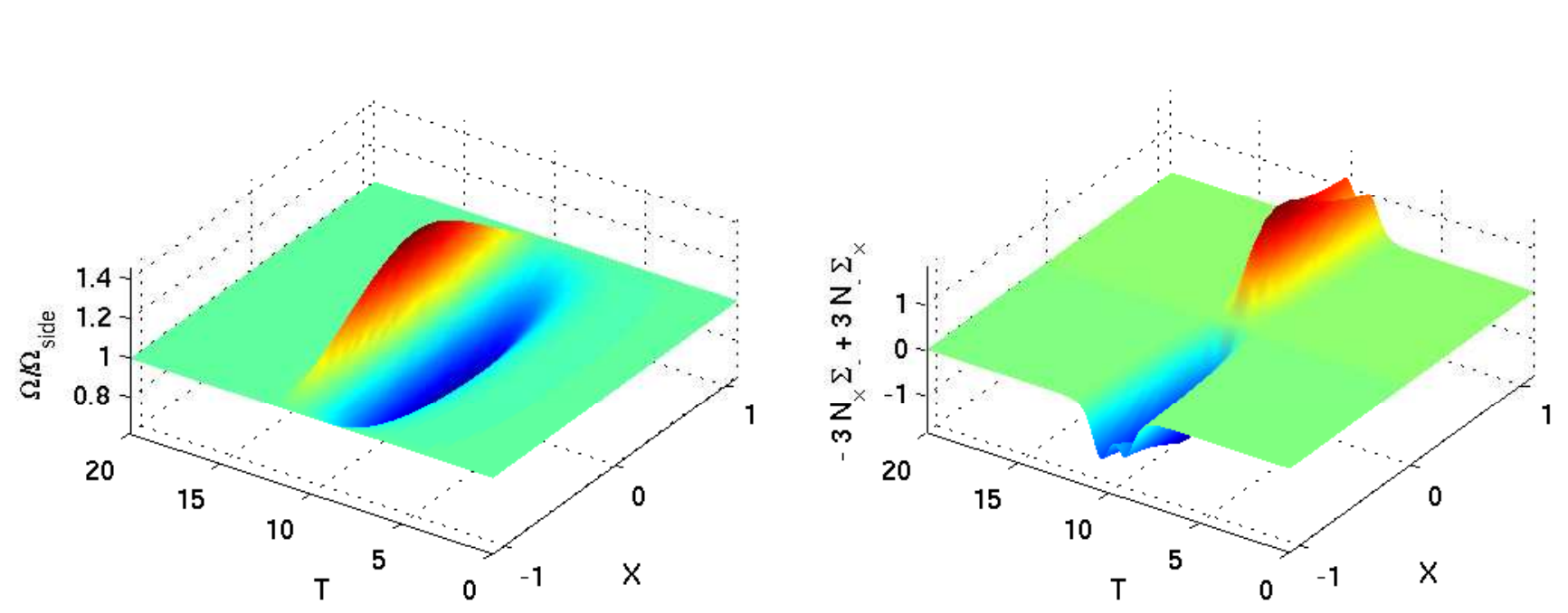}
    \caption{(Left panel) Corresponding ratio $\Omega/\Omega_{X=-1.1}$. $\Omega$ develops first an underdensity and then an overdensity towards singularity. (Right panel) The expression $-3\Nc\Sm+3\Nm\Sc$ that drives $v$.}
    \label{g_rad_v3-sideo}     
\end{center}
\end{figure}

Zooming in at $X_0=0$, at first $v$ has negative slope, and this slows down the
evolution of $v$ towards $\pm 1$.  When the spike transition occurs, the expression
$-3\Nc\Sm+3\Nm\Sc$ in the $r$ term becomes large and drives $v$ in the opposite
direction so much so that the slope of $v$ becomes positive.  Interpreting this
result in the expanding direction, we say that this spike transition alters the
slope of $v$ and hence alters the sign of the divergence term from positive to
negative.  See Figures~\ref{g_rad_v3} and~\ref{g_rad_v3-sideo}.  Interptreting
Figure~\ref{g_rad_v3-sideo} towards the singularity; $\Omega$ first develops an
underdensity, then the expression $-3\Nc\Sm+3\Nm\Sc$ becomes large and changes the
sign of the divergence term from positive to negative, $\Omega$ now develops an
overdensity, and lastly the spike transition takes the solution to the Kasner arc
where $v$ is stable and the divergence term becomes small.  The initial and final
flatness in the plot of $\Omega/\Omega_\text{side}$ is an artifact of the initial
data at $T=0$ and the fact that the scale of overdensity became super-horizon scale
from $T=12$ onwards (because the particle horizon is shrinking).  Interpreting
Figure~\ref{g_rad_v3-sideo} forward in time, the overdensity becomes an underdensity
at around $T=10$, which then becomes flat (but only because 
of the initial data we prescribed) at $T=0$.  The large expression $-3\Nc\Sm+3\Nm\Sc$ at around $T=10$
alters the sign of the divergence term from positive to negative.  The negative
divergence term undoes the underdensity.

In this case $\Omega$ became essentially homogeneous towards the singularity at 
subhorizon scale.
We saw this in \cite{art:ColeyLim2012} but gave an incorrect interpretation. We incorrectly attributed the primary role of inhomogeneity generation to the spike transition, while it should have been attributed to the sign change in 
the tilt. Spike transitions play the secondary role of driving the sign of the divergence term towards negative.

Note that although a spike has a "handedness" to it, the expression $-3\Nc\Sm+3\Nm\Sc$ does not. 
This can be seen by running a similar simulation with $w = -1.5$. This changes the sign of the triggers $\Nm$ and $\Sc$, but $\Nc$ and $\Sm$ maintain their 
signs. From the expression for $r$ we see that $r$ maintains its sign as a result ($r$ is positive for $X>0$). So this spike with $w=-1.5$ still does the same thing to $v$ and the divergence term.

We can conclude that, if $v$ is small (unlike in the previous section where $v^2$ is close to 1), then the spike transition manages to drive the divergence term towards negative in the expanding direction.
While the divergence term itself has no preference to be positive or negative, spike transitions drive the divergence term towards negative whenever they intersect a sign change of the tilt.
In a general model without symmetry, this intersection occurs along curves, and generates local overdensity in the matter along these curves, leading to a web of local overdensity.

\section{Conclusion}

In \cite{art:ColeyLim2012} we studied orthogonally transitive $G_2$ models with a tilted radiation fluid.
We found that spikes, which are purely gravitational inhomogeneities, 
leave imprints on the radiation fluid in the form of local overdensities.
We have obtained further numerical evidence for the existence of GR matter perurbations, which
support the results of \cite{art:ColeyLim2012}.
However, what we did not realize was that this was partly due to the non-negligible 
divergence term caused by the instability in the tilt.
In this paper we look at the role of the tilt more carefully.
We conclude that tilt instability plays the primary role in the formation of matter inhomogeneities. 
A negative divergence term creates a local overdensity.
While the divergence term itself does not prefer one sign or another, 
spikes drive the divergence term towards negative, creating a web of local overdensities.

We have only explored the orthogonally transitive $G_2$ case, which has one tilt degree of freedom.
In general there are three tilt degrees of freedom, whose dynamics is even richer.
One can imagine that the local overdensities are compressed anisotropically in three different directions, 
leading to generally elliptical lumps of local overdensities, distributed along surfaces, and with higher densities
along intersections of these surfaces.
In addition to spikes and large divergence terms, we also encountered shock waves in this paper, 
but we know almost nothing about the role of these shock waves.
Recall that in \cite{art:ColeyLim2012} we discussed incomplete spikes, which have not been studied here. 
Incompletely spikes, if they terminate before they are halfway done, generate significant local underdensities,
provided that they coincide with a non-negligible divergence term.

\section{Discussion}

In \cite{art:ColeyLim2012}
we explicitly showed   that in GR spikes naturally occur in a class of
$G_2$ models leading to inhomogeneities that, due to
gravitational instability, leave small residual imprints on matter in the
form of matter perturbations.
We have shown that the tilt provides another mechanism in generating matter inhomogeneities through
the divergence term. In fact, at least in the  models we have studied,
the tilt plays the primary role in generating matter inhomogeneities, and
both overdensities and underdensities are generated.

We are particularly interested in recurring and complete distributed
spikes formed in the oscillatory regime (or recurring spikes for short), and their
imprint on matter and structure formation.
The residual matter overdensities from
recurring spikes are not local but form on surfaces.
In particular, in the $G_2$ models the
inhomogeneities can occur on a surface, and in general spacetimes the inhomogeneities can
occur along a line, leading to matter inhomogeneities forming on walls or surfaces.
Indeed, there are tantalising hints (from dynamical and numerical analyses) that
filamentary structures and voids would occur naturally in this scenario.

We have speculated  \cite{art:ColeyLim2012} as to whether these
recurring spikes might be an alternative to the inflationary mechanism for
generating matter perturbations and thus act as seeds for the subsequent formation of large scale structure.
Superficially, at least, there are some
similarities with perturbations and structure formation created in cosmic string models.
The inhomogeneities occur on closed circles or infinite lines \cite{art:ColeyLim2012}, 
similar to what happens in the case of
topological defects,
and it is expected that a mechanism akin to the Kibble causality mechanicsm will
ensure that ``defects'' form and persist to the present time.

Topological defects, such as domain walls, cosmic strings and global textures,
are generically produced during phase transitions in the very early Universe
in the framework of (supersymmetric) grand unified theories \cite{art:Brandenberger1994,book:VilenkinShellard1994,art:DuPlessisBrandenberger2013}.
The reason why topological defects can play a role in structure formation in
the early Universe is because they carry energy which
leads to an extra attractive gravitational force, and hence the defects can act as
seeds for cosmic structures.
In particular, cosmic strings are one-dimensional topological defects which
lead to a scale-invariant spectrum
of cosmological perturbations (and, among other things,
non-Gaussianities)
\cite{art:Brandenberger1994,book:VilenkinShellard1994,art:DuPlessisBrandenberger2013}.
Cosmic strings contribute to the power
spectrum of density fluctuations
which affect cosmic microwave background (CMB) radiation  temperature anisotropy maps  \cite{art:Dunkleyetal2009,art:Ade2013,art:Urrestilla2011}.
Although it was shown
that cosmic strings (for example)
can be ruled out
as the {\em{unique}} source of density perturbations leading to the observed structure
formation \cite{art:Sakellariadou2006,art:Hindmarsh2011}, when cosmic strings are included within
inflationary models the tension with
WMAP and PLANCK data can be reduced \cite{art:Dunkleyetal2009,art:Ade2013,art:Urrestilla2011}.

We are also interested in incomplete spikes.
Eventually, the oscillatory regime ends when $\Omega$ is no longer negligible, and
some of the spikes  are in the middle of transitioning, leaving
inhomogeneous imprints on the matter result.
The residuals from an incomplete spike
might, in principle, be large and thus affect structure formation.
Indeed, the numerics suggest
$\Omega$ develops a void at a spike location \cite{art:ColeyLim2012}.

The incomplete spikes associated with Kasner saddle points occur generically in the early universe.
Saddles, related to Kasner solutions and FLRW models, may also occur at late times, and may also
cause spikes/tilt that might lead to further matter inhomogeneities, albeit non-generically.
In particular, since the flat FLRW solution appears to have a 3-dimensional unstable manifold,
a spike can still form (towards the future) around a point.\footnote{
The open FLRW solution appears to be stable, so no spikes can form. Along general world lines
one has an open FLRW-Milne-void solution, but around isolated points one
gets flat FLRW towards the future.}
This kind of spike is potentially very interesting from the physical point of view.
Further investigation is needed to confirm these speculations.

Both the incomplete spikes and the late time non-generic inhomogeneous spikes
 might lead to the existence of
exceptional structures on large scales. 
In the standard cosmological model spatial homogeneity is only valid
on scales larger than 100-115 Mpc
\cite{art:Scrimgeouretal2012,art:Yadav2010}
(and only then in some statistical sense when averaged
on large scales
\cite{art:Wiltshire2011,art:Coley2010,art:Coleyetal2005,art:Coley2010b,art:Buchert2008}).
However, in the actual Universe the distribution of matter is far from homogeneous
on scales less than 150-300 Mpc.
There are a number of very large structures, such as 
the Shapley supercluster \cite{art:Tullyetal1992}, 
the Sloan great wall  \cite{art:Gottetal2005}, the CfA (great) wall 
\cite{art:ShethDiaferio2011} and a number of very
large quasar groups (LQGs) 
\cite{art:Clowes2012} and some enormous local voids  \cite{art:Rizzi2007,art:Frith2003}
and void complexes on scales up
to 450 Mpc \cite{art:Park2012}.

Such large inhomogeneities in the distribution
of superclusters and voids on scales of 200-300 Mpc and above, and
especially the ``Huge-LQG'' (with characteristic volume of size $\sim$ 500 Mpc)
and its proximity to the CCLQG at the
same redshift  ($\sim 1.27$) \cite{art:Clowes2012},
implies that the Universe is
perhaps not homogeneous
on these scales, and
are potentially in conflict
with  the cosmological principle and the standard concordance ($\Lambda$CDM) cosmological model
\cite{art:ShethDiaferio2011,art:Park2012}).
An alternative GR  spike mechanism for naturally generating (a small number of)
exceptional structures at late times (in additional to the usual
distribution of structures produced in the standard model) may resolve some tension with
cosmological observations.

Finally,
in \cite{art:NungesserLim2013}
the existing relationship between polarized
Electrogowdy spacetimes and Gowdy spacetimes was exploited to find explicit solutions
for electromagnetic spikes.
New inhomogeneous
solutions, called the electric and the magnetic spike solutions, were presented.
It will be interesting to see how gravitational
spikes, electromagnetic spikes and perfect fluid interact to generate
inhomogeneities in the fluid density and the electromagnetic field.  This
will provide a relativistic (non-Newtonian and non-quantum) mechanism to
produce primodial galaxies and intergalactic magnetic fields, and explain
the web-like distribution of these structures.
For example, cosmic strings can be responsible for the production of highly
energetic bursts of particles, and they can help seed coherent magnetic fields on galactic scales \cite{art:BrandenbergerZhang1999}.
We will return to this question in the future.

\section*{Acknowledgement}
This work was supported, in part, by NSERC of Canada. We thank R. H. Brandenberger and the Department of Physics and Astronomy, University of Canterbury for helpful discussions.

\bibliography{ecites}

\end{document}